\documentclass[preprint,showpacs,amsmath,amssymb,pra]{revtex4}
\usepackage{dcolumn} % Align table columns on decimal point
\usepackage{bm}      % bold math
\usepackage[dvips]{graphics}
\newcommand{\wek}[1]{\vec{\bf {#1}}}
\newcommand{\pdt}{\frac{\partial}{\partial t}}
\newcommand{\ckk}[1]{{\cal K}_{#1}(\wek{v}\leftarrow\wek{v}\,')}
\newcommand{\xket}[1]{\left| \, #1 \, \right\rangle}
\newcommand{\xbra}[1]{\left\langle \, #1 \, \right|}

\begin{document}
%%%-----------------------------------------------------------------
\title{On the positivity of Bloch--Boltzmann equations}
\author{Stanis{\l}aw Kryszewski }
\email{fizsk@univ.gda.pl}
\author{Justyna Czechowska}
\affiliation{Institute of Theoretical Physics and Astrophysics,
             University of Gda{\'n}sk, ul. Wita Stwosza 57,
             80-952 Gda{\'n}sk, Poland}
\date{\today}
%%%-----------------------------------------------------------------
\begin{abstract}
In a large variety of spectroscopical applications Bloch-Boltzmann
equations (BBE) play an essential role. They describe the evolution
of the reduced density operator of an active atom which is coupled
to radiation (Bloch part) and which interacts collisionally with
the perturber gas (Boltzmann part). The standard approach to the
collisional part is well-known from the literature. It preserves
hermiticity and normalization, but the question whether it
preserves positivity seems to remain open. The completely
positive BBE were recently derived via the general master
equation techniques. These two approaches are applied for
a model of $n$-level nondegenerate atom. We show that within
this model both approaches to the collisional part of BBE are
equivalent -- give the same physical predictions. The approach
based upon master equation techniques guarantees the preservation
of hermiticity, normalization and positivity. The proven
equivalence ascertains that the standard approach also preserves
positivity. Moreover, some aspects of the standard derivation
(which atomic states do contribute to the evolution)
are clarified by the established equivalence.
\end{abstract}
                              %display  desired
\maketitle
%%%-----------------------------------------------------------------

%%%-----------------------------------------------------------------
\section{Introduction}

One of typical, spectroscopically important experimental situation
occurs when a gaseous mixture is irradiated by the external light
source. The mixture consists of active atoms which couple to the
incident radiation and of a usually much denser inert gas.
The perturbers atoms collide with the active ones, thereby
influencing their behavior. Then, many properties of such
system can be studied, both experimentally and theoretically.
The literature devoted to such problems is large, so we
indicate only some essential monographs \cite{demt,rash,scul,coh}.

Usually, only the active atoms are of interest, hence they
must be theoretically described within the density operator
formalism. The equations of motion for the active-atom density
operator may be called, in the absence of any better name,
Bloch-Boltzmann equations (BBE). The Bloch part describes the
interaction between the active atoms and radiation. It is a
generalization of the well-known two-level optical Bloch
equations to a more general multilevel case. The collisional
interaction between active atoms and perturbers is given by
suitably constructed collision integrals. This contribution to
equations of motion might be called the Boltzmann part.

The recent advances in the fundamentals of quantum mechanics
and in the quantum information theory have shown the
importance of the preservation of the basic properties
of any density operator: hermiticity, normalization and
positivity (for an excellent review, see \cite{keyl} and the
references given therein). These essential properties of
the density operator must be preserved by any theoretical
formalism. The aim of this work is to discuss this point
for the spectroscopically important situation which was
sketched above. Since radiative (or Bloch) part of the
corresponding equations of motion is already well
investigated we shall focus our attention on the
collisional (Boltzmann) part.

In the recent paper \cite{ak} a gaseous mixture of two
species: $A$ -- active atoms and $P$ -- perturbers,
was considered. The densities of these two components are assumed
to satisfy the relation: $N_{A} \ll N_{P}$, and an equation of
motion for the density operator of the $A$-atom interacting
collisionally with the perturbers, i.e., is rederived.
This  approach is based upon general master equation
(ME) techniques (for a review, see \cite{al}) in the
spirit of Lindblad-Gorini-Kossakowski-Sudarshan method.
The employed technique ensures that the $A$-atom density
operator possesses all the necessary properties: it is
hermitian, normalized and positive-definite for all instants
of time. It is perhaps worth mentioning that the Bloch part
(i.e., radiative one) of the equations of motion for the
active-atom density operator is usually derived within
ME techniques (see \cite{coh}). Therefore, this contribution
to BBE is certain to preserve the mentioned properties of
the density operator. This is also the reason why we restrict
our attention to the collisional part of BBE.

The other approach to the derivation of the Boltzmann part
of the BBE is known since the pioneering work of Snider
\cite{snid}. Then, it was refined by other authors
and employed in a variety of practical applications (see, for
example \cite{berm01,hube}). The monograph \cite{rash}
by Rautian and Shalagin seems to give the most comprehensive
review of the derivation of collision integrals appearing in BBE,
especially for specific, spectroscopy-oriented applications.
In the following, we shall call their presentation the standard
approach to the derivation of BBE. It is, perhaps, worth noting
that the derivation given by Rautian and Shalagin leads to the
appearance of some Kronecker-like delta factors which perform
the role of "state selectors" -- select the atomic states
which contribute to the evolution of the A-atom density
operator. The authors themselves say that their procedure
is open to question (\cite{rash}, p.42).

Working within the standard approach it is relatively
straightforward to prove that it preserves hermiticity of the
atomic density operator. The proof that the normalization
is also preserved requires one to invoke the optical theorem
of the quantum multichannel scattering theory. However,
we are not aware of any studies in which the preservation of
positivity is investigated. Discussion of this point is the
main aim of our work. The secondary aim of our work is to
investigate the validity od the "state selection" mechanism
proposed by Rautian and Shalagin

It is not our aim to present the details of the derivations,
or the underlying physical and mathematical assumptions,
of the two indicated approaches to Boltzmann part of BBE.
These aspects are well-documented in the literature
\cite{rash,ak}, so it seems to be no need to repeat them here.
We shall only use the results relevant to the main subject
of our discussion.

Sec. II is devoted to brief presentation of the collisional
contributions to BBE obtained within the master equation
approach and in the standard one. We do not derive them, but
simply state the results -- the results which are essential
for further discussion.

In the next section we adapt the general formulas of the
previous section to the model of $n$-level atom with
nondegenerate energies. We construct the collisional parts
of BBE corresponding to both approaches. In the last
subsection we argue that the obtained results are, in fact,
equivalent. This allows us to state that the standard approach
(within the adopted model) preserves the positivity of the
active-atom density operator. The proof of this fact constitutes
the main result of this work.

Finally, the fourth section contains some remarks which may
be useful for some further research, that is for discussing
the preservation of positivity for more general models studied
within the standard approach (master equation approach is
guaranteed to do so). Moreover, we hope that some of our
remarks will be useful to provide the standard approach
with some more rigorous standing. Namely, the equivalence
of both approaches validates the the Kronecker delta-like
"state selective" factors which appear in the standard approach
of Rautian and Shalagin.

%%%-----------------------------------------------------------------
\section{Two approaches to derivation of the collisional part
of BBE}

%%%-----------------------------------------------------------------
\subsection{Master equation approach \label{sec:me}}

The derivation of the Boltzmann part of BBE via the master
equation technique is given in the recent paper \cite{ak},
where the necessary assumptions are also discussed.
This is a mathematically rigorous, although fairly involved method.
It is not our purpose to obscure the physical discussion by
mathematical technicalities, therefore, we will present here
only the most essential results of the theory given in \cite{ak}.

In order to ensure the preservation of hermiticity, normalization
and positivity of the reduced atomic density operator,
the corresponding equation of motion must be of the following
general (so-called Lindblad-Kossakowski-Gorini)
form (see also \cite{al})
\begin{align}
   \pdt \: \rho_{\alpha}
   = - \frac{i}{\hbar}
       \bigl[ H_{\alpha},\;\rho_{\alpha} \bigr]
   & + \sum_{\beta} \sum_{\xi}
       \hat{S}_{\alpha \beta}^{\,\xi} ~\rho_{\beta}
       (\hat{S}_{\beta \alpha}^{\,\xi} )^{\dagger}
\nonumber \\
   & - \frac{1}{2} ( \hat{B}_{\alpha}\rho_{\alpha}
         + \rho_{\alpha} \hat{B}_{\alpha} ),
\label{sjj1} \end{align}
where $\rho_{\alpha}$ is a (reduced) density operator of
an $A$-atom. It is parameterized by an index $\alpha$ which,
under the suitable additional assumptions \cite{ak}, can be
shown to correspond to the velocity of a considered atom.
Let ${\cal H}_{\alpha}$ denote a Hilbert space of the
atomic states. Then  the quantities appearing in
Eq. \eqref{sjj1} are defined as mappings (operators):
\begin{subequations} \label{sjj2}
\begin{align}
   & H_{\alpha} = H_{\alpha}^{\dagger}
     : {\cal H}_{\alpha} \rightarrow {\cal H}_{\alpha},
     ~~\mathrm{(Hamiltonian)},
\label{sjj2a} \\
   & \hat{S}_{\alpha \beta}^{\,\xi}
     : {\cal H}_{\beta} \rightarrow {\cal H}_{\alpha},
\label{sjj2b} \\
   & (\hat{S}_{\beta \alpha}^{\,\xi})^{\dagger}
     : {\cal H}_{\alpha} \rightarrow {\cal H}_{\beta},
     ~~ \bigl( \mathrm{H.C. ~of} ~\hat{S}_{\alpha \beta}^{\,\xi}
        \bigr),
\label{sjj2c} \\
   & \hat{B}_{\alpha}
     = \sum_{\xi} \sum_{\beta}
       (\hat{S}_{\alpha \beta}^{\,\xi})^{\dagger}
       \hat{S}_{\beta \alpha}^{\,\xi},
\label{sjj2d} \end{align}
\end{subequations}
The general master equation \eqref{sjj1} is then adapted to
our needs -- to describe the active-atom-perturber collisional
interaction. These steps are also discussed in \cite{ak},
so we only state the results essential to further discussion.

We consider a $n$-level A-atom immersed in much denser perturbers.
The perturbers (assumed to be structureless particles) thermalize
very rapidly, hence their distribution is simply Maxwellian
\begin{equation}
  W^{(P)}(\wek{v})
  ~=~ \left( \frac{1}{\pi u_{p}^2} \right)^{3/2} \:
      \exp \left( - \frac{\wek{v}^{\:2}}{u_{p}^2} \right),
\label{sjmax} \end{equation}
with $u_{p}^{2}=2k_{B}T/m_{p}$, where $m_{p}$ is the mass of
the perturber atom.

Let us now take the Hamiltonian of the free A-atom as
\begin{equation}
   H_{A} = \sum_{k}^{n} \hbar \omega_{k} \;
           \xket{k}\xbra{k}
\label{sjaham} \end{equation}
where the eigenfrequencies $\omega_{k}$ may, in general,
be degenerate.

Next, let  $\{ S_{a} \}$ be a basis in the space of operators
acting on the Hilbert space of A-atom states $\{ \xket{k} \}$.
These operators satisfy the relation
\begin{equation}
   \bigl[ H_{A}, ~S_{a} \bigr]
   = \hbar \Omega_{a} S_{a},
     \hspace*{5mm} a = 1,2, \ldots\ldots,n^{2},
\label{sjsb} \end{equation}
where the quantities $\Omega_{a}$ are identified as Bohr
frequencies.

Within this framework, the collisional part of the master
equation becomes \cite{ak}:
\begin{align}
   \pdt & \rho(\wek{v}) \Bigl|_{coll.}
   = - \frac{1}{2} \sum_{a,b}  \gamma_{ba}(\wek{v})
       \Bigl[ S_{a}^{\dagger} S_{b}, ~\rho(\wek{v}) \Bigr]_{(+)}
\nonumber \\
   & + \sum_{a,b} \int d\wek{v}\,'
       ~\ckk{ab} S_{a} \rho(\wek{v}\,') S_{b}^{\dagger},
\label{sjme} \end{align}
where the $(+)$ subscript denotes the anticommutator, and
$\rho(\wek{v}) = \rho(\wek{r},\wek{v},t)$
is the reduced density operator of an A-atom with respect
to internal variables (states) but a phase-space distribution
with respect to position and velocity.
The relaxation (collisional) rate $\gamma_{ba}(\wek{v})$
is specified as
\begin{equation}
    \gamma_{ba} \equiv  \gamma_{ba}(\wek{v})
    = \int d\wek{v}\,'~{\cal K}_{ba} (\wek{v}\,'
      \leftarrow \wek{v}).
\label{sjg1} \end{equation}
Finally, it can be shown \cite{ak,al} that the matrix
$\ckk{ab}$ is expressed as
\begin{align}
   {\cal K}_{ab}&(\wek{v} \leftarrow \wek{v}\,') =
\nonumber \\
   &= 2 N_{P} \; \delta_{\Omega_{a},\Omega_{b}}
       \int d\wek{v}_{r1} \int d\wek{v}_{r}
       ~W^{(P)}(\wek{v}\,' - \wek{v}_{r1})
\nonumber \\
   &\times \delta^3 \left[ \wek{v} - \wek{v}\,'
        -\frac{\mu}{m_a} \left(\wek{v}_{r} - \wek{v}_{r1} \right)
                \right]
\nonumber \\
& \times \delta \left( v_{r}^{2} - v_{r1}^{2}
      + \frac{2\hbar \Omega_{a}}{\mu} \right)
\nonumber \\
& \times f_{a}( \wek{v}_{r} \leftarrow \wek{v}_{r1}) \:
         f_{b}^{\ast}(\wek{v}_{r} \gets \wek{v}_{r1}).
\label{sjk1} \end{align}
The employed notation is as follows. $m_{a}$ is the mass of an
A-atom, while $\mu$ is the reduced mass of A-P colliding partners.
$\wek{v}_{r}$ and $\wek{v}_{r1}$ are the relative velocities.
$N_{P}$ is the density of the perturber gas.
The functions ${f}_{a}(\wek{v}_{r} \leftarrow %
\wek{v}_{r1})$ are connected with usual (taken in the
center-of-mass frame) scattering amplitudes:
\begin{equation}
   \sum_{a}
   f_{a}(\wek{v}_{r} \gets \wek{v}_{r1}) S_{a} =
   \sum_{j,k=1}^{n}
   f (j,\wek{v}_{r} \gets k,\wek{v}_{r1})
   \left| j \right \rangle  \left \langle k \right|.
\label{sjff1} \end{equation}
We note that the $\ckk{ab}$ matrix is clearly hermitian and
positive definite. Hermiticity of the matrix $\ckk{ab}$ implies
that $\gamma_{ba}^{\ast}(\wek{v}) =  \gamma_{ab}(\wek{v})$.
These ensures the preservation of hermiticity of the A-atom
reduced density operator.

The factor $\delta_{\Omega_{a},\Omega_{b}}$ in Eq. \eqref{sjk1}
has the sense of the Kronecker delta
\begin{equation}
   \delta_{\Omega_{a},\Omega_{b}}
   =  \left\{
      \begin{array}{cc}
      0 \hspace*{10mm} \mathrm{for}
        \hspace*{10mm} \Omega_{a} \ne \Omega_{b}, \\
      1 \hspace*{10mm} \mathrm{for}
        \hspace*{10mm} \Omega_{a} = \Omega_{b}
  \end{array}
 \right.
\label{sjod} \end{equation}
This factor appears due to the secular approximation which is
necessary in derivation of master equation (as is clearly
shown in \cite{coh}). The given $\ckk{ab}$ matrix also ensures
the momentum and energy conservation.

We stress that the resulting collisional ME \eqref{sjme} preserves
all the necessary properties of the A-atom reduced density
operator $\rho(\wek{v})$. Preservation of hermiticity follows
from hermiticity of $\ckk{ab}$ matrix. Next, relation \eqref{sjg1}
ensures that
\begin{equation}
    \pdt \int d\wek{v} ~{\rm{Tr}} \{ \rho( {\wek{v}}) \} = 0,
\label{norm} \end{equation}
as necessary for preservation of normalization. Finally,
preservation of positivity is ensured by general Lindblad
structure of the master equation \eqref{sjme}.

%%%-----------------------------------------------------------------
\subsection{Standard approach \label{sec:ss}}

The standard derivation of the Boltzmann part of BBE as presented
by Rautian and Shalagin \cite{rash} is rather lengthy and fairly
complicated. It is based upon two physical assumptions:
$N_{A} \ll N_{P}$, so that the A-P collisions are binary,
and the duration of the collision is by far the shortest time
scale, so that the impact approximation is valid
(see \cite{rash}, p.31). The general von Neumann equation
for the density operator for the entire system is truncated
(traced) to an equation for a single A-atom. The interaction
with the perturbers is then considered within the framework
of the time-dependent scattering theory. The collision
integrals are then expressed (similarly as in \cite{snid})
in terms of the elements of the scattering $T$-matrix,
which are subsequently reexpressed by usual scattering
amplitudes. Further steps consist in semiclassical approximation
which leads to the Boltzmann terms of the following shape
\begin{align}
  & \pdt \: {\rho_{\alpha \alpha^{\:'}}}(\wek{v}) \Big|_{coll.}
  = - \sum_{\alpha_1 \alpha_1^{\:'}}
    ~\Gamma(\alpha \alpha^{\:'},\wek{v} \: \big| \:
        \alpha_1 \alpha^{\:'}_1)
    \rho_{\alpha_{1} \alpha^{\:'}_{1}}(\wek{v})
\nonumber \\
& + \sum_{\alpha_{1} \alpha_{1}^{\:'}} \int d\wek{v}_{1}
   ~\mathcal{K} (\alpha \alpha^{'}, \wek{v} \: \big| \:
   \alpha_{1}\alpha_{1}^{'}, \wek{v}_{1} \bigr)
   \rho_{\alpha_{1} \alpha^{\:'}_{1}}(\wek{v}_{1}),
\label{sjse} \end{align}
where $\rho_{\alpha \alpha^{\:'}}(\wek{v})$ denotes the matrix
elements  of the A-atom reduced density operator (which has
the same sense as in the ME approach). The indices $\alpha$
should be understood as multiple ones (atomic states may be
labelled by several quantum numbers).

The collision rate appearing in the first term is given as
\begin{align}
  \Gamma & \bigl( \alpha \alpha^{'}, \: \wek{v} \: \bigl| \:
  \alpha_{1} \alpha_{1}^{'} \bigr) =
\nonumber \\
&    = N_{P} \left( \frac{2 \pi \hbar}{i \: \mu} \right)
       \int d\wek{v}_{r} ~W^{(P)} (\wek{v}-\wek{v}_{r})
\nonumber \\
&   \times \Bigl[
    f( \alpha, \wek{v}_{r} \leftarrow
       \alpha_{1}, \: \wek{v}_{r} \bigr)
   \delta( \omega_{\alpha \alpha_{1}})
   \delta_{\alpha^{'} \alpha_{1}^{'}}
\nonumber \\
& \hspace*{5mm}
  - f^{*}( \alpha^{'}, \: \wek{v}_{r} \leftarrow
           \alpha_{1}^{'}, \: \wek{v}_{r} \bigr)
   \delta( \omega_{\alpha^{'} \alpha_{1}^{'}})
   \delta_{\alpha \alpha_{1}}
    \Bigr].
\label{sjsgam} \end{align}
We note that this collisional rate is given by the elastic
forward scattering amplitudes. $\delta_{\alpha \alpha_{1}}$
are simple  Kronecker-type deltas, while the factors
$\delta( \omega_{\alpha \alpha_{1}}) =%
\delta( \omega_{\alpha} - \omega_{\alpha_{1}})$ have meaning
similar to that defined in Eq. \eqref{sjod}. These delta
factors ensure energy conservation. Their origin and significance
will be discussed later.

The second term of Eq. \eqref{sjse} contains the collision
kernel specified as
\begin{align}
    \mathcal{K}
  & (\alpha \alpha^{\,'},\wek{v} \: \bigl| \:
     \alpha_{1}\alpha_{1}^{\,'}, \wek{v}_{1} ) =
\nonumber \\
  & = \bigl\{ \delta( \omega_{\alpha \alpha^{'}})
     \delta( \omega_{\alpha_{1} \alpha_{1}^{'}})
\nonumber \\
  & \hspace*{12mm}
  + \bigl[ 1 - \delta( \omega_{\alpha \alpha^{'}}) \bigr]
    \delta( \omega_{\alpha \alpha_{1}})
    \delta( \omega_{\alpha^{'} \alpha_{1}^{'}}) \bigr\}
\nonumber \\
  & \times 2 N_{P} \int d \wek{v}_{r} \int d\wek{v}_{r1}
      ~W^{(P)} ( \wek{v}_{1} - \wek{v}_{r1} )
\nonumber \\
  & \times \delta \left( \wek{v} - \wek{v}_{1}
    - \frac{\mu}{m_{a}} (\wek{v}_{r} - \wek{v}_{r1} ) \right)
\nonumber \\
  & \times
    f(\alpha, \wek{v}_{r} \leftarrow
      \alpha_{1}, \wek{v}_{r1} )
   f^{\ast}( \alpha^{'}, \wek{v}_{r} \leftarrow
             \alpha_{1}^{'},\wek{v}_{r1} )
\nonumber \\
& \times \delta \left( v_{r}^{2} - v_{r1}^{2}
   + \frac{2}{\mu} ( E_{\alpha} - E_{\alpha_{1}} ) \right).
\label{sjsker} \end{align}

The peculiar feature of the Rautian and Shalagin derivation
consists in the appearance of the Kronecker-type delta
factors. These factors play a selective role, indicating that
not all matrix elements $\rho_{\alpha_{1} \alpha^{\:'}_{1}}$
(in the rhs of Eq. \eqref{sjse}) contribute to the evolution
of $\rho_{\alpha \alpha^{\:'}}$.
The origin of these factors is explained by Rautian and Shalagin
in the following way.

The $T$-matrix elements $\langle\,\alpha,\wek{p}_{r}\,%
|\,\hat{T}\,|\,\alpha_{1},\wek{p}_{r1}\,\rangle$ (taken in
the center-of-mass frame) include phase factors of the type of
\begin{equation}
 \exp \left[ \, \frac{i t}{\hbar}
      \left( \, E_{\alpha} - E_{\alpha_1} \, \right) \, \right].
\label{sjexp} \end{equation}
When Bohr frequencies $\omega_{\alpha \alpha_1} =%
(E_{\alpha} - E_{\alpha_1})/\hbar$ are non-zero, the
corresponding exponentials oscillate rapidly and their
contribution to the overall evolution averages out virtually
to zero. In other words, only those states for which Bohr
frequencies are close to zero contribute significantly
to collision integrals. This argument of Rautian and Shalagin
gives rise to the $\delta( \omega_{\alpha \alpha_{1}})$-type
factors in the collision rate and kernel. It must be, however,
stressed that these simple delta-like terms appear due to the
assumption that the perturbers are unpolarized.
If this assumption does not hold, the structure of the
corresponding delta-like factors would be different and
more complicated since these factors would include Bohr
frequencies also for perturbers.

Rautian and Shalagin discuss the role of the exponential factors
but do not carry their calculation as far as we did \cite{jczech}.
They retain factors \eqref{sjexp} in their formulas and
comment only verbally on their significance. Moreover,
up to our knowledge, no other authors conduct such
a discussion. The reason, perhaps, is that even Rautian and
Shalagin doubt upon the validity of such a "selective
mechanism" expressed by the exponential phase factors
\eqref{sjexp} and consequently by delta-type factors
(see their discussion at the top of p. 42 in \cite{rash}).
The significance of the discussed delta-like factors reduces to,
roughly speaking, that {\em "like is excited by like"}
(\cite{rash}, p. 42). This means that the equation of motion
\eqref{sjse} connects populations with populations and coherences
with coherences.

In the forthcoming we shall return to the discussion of this point,
when we shall compare the ME results with the standard
ones for a more specific model of an active atom.
This will allow us to shed some new light onto the role played
by the "selective" delta-like factors.

Furthermore, we note that it is relatively easy to show that
\begin{equation}
  \Gamma^{\ast}
  \bigl( \alpha \alpha^{'}, \: \wek{v} \: \bigl| \:
         \alpha_{1} \alpha_{1}^{'} \bigr)
  = \Gamma \bigl( \alpha^{'} \alpha, \: \wek{v} \: \bigl| \:
  \alpha_{1}^{'} \alpha_{1} \bigr),
\label{sjgast} \end{equation}
and similarly
\begin{equation}
    \mathcal{K}^{\ast}
    (\alpha \alpha^{\,'},\wek{v} \: \bigl| \:
     \alpha_{1}\alpha_{1}^{\,'}, \wek{v}_{1} )
    = \mathcal{K}(\alpha^{\,'} \alpha,\wek{v} \: \bigl| \:
      \alpha_{1}^{\,'}\alpha_{1}, \wek{v}_{1} ).
\label{sjkast} \end{equation}
Both these relations ensure that the evolution given by
Eq. \eqref{sjse} preserves the hermiticity of the A-atom
density operator.

The preservation of the proper normalization of the density
operator on one hand follows directly from the general formalism
employed by Rautian and Shalagin. On the other hand, it should
be also possible to prove that the summation and integration
over $\wek{v}$ of the diagonal equations \eqref{sjse}
yields zero, as required by normalization. The complicated
structures of the rate $\Gamma$ and kernel $\mathcal{K}$
make it a rather formidable task (at least in general).
It seems that it is better to use the general equations
for some specific model of $A$-atom structure. Then, checking
that normalization is indeed preserved should be much easier.
We shall do so in further section. It is, however, not clear
whether the positivity of $\rho(\wek{r},\wek{v},t)$ is also
preserved. There seems to be no compelling, rigorous argument
to state so for certain. We shall later return to the discussion
of this very important issue.

Finally, we note that the collisional rate \eqref{sjsgam} has
the following interesting property
\begin{align}
  \mathrm{Re} & \left[
     \Gamma \bigl( \alpha \alpha^{'}, \: \wek{v} \: \bigl| \:
        \alpha \alpha^{'} \bigr)
     \right] =
\nonumber \\
  &  = \frac{1}{2} \left[
    \Gamma \bigl( \alpha \alpha, \: \wek{v} \: \bigl| \:
                  \alpha \alpha \bigr)
   + \Gamma\bigl( \alpha^{'} \alpha^{'}, \: \wek{v} \: \bigl| \:
        \alpha^{'} \alpha^{'} \bigr)
   \right],
\label{sjgpro} \end{align}
which will be useful in the further discussion.

%%%-----------------------------------------------------------------
\section{Discussion \label{sec:dd}}

%%%-----------------------------------------------------------------
\subsection{General comments and atomic model}

Equations \eqref{sjme} and  \eqref{sjse} representing two
approaches to the QMBE are of the similar,
although not necessarily identical form. These external
dissimilarities have led us (see \cite{ak}) to the supposition
that the standard approach may not preserve
the positive definiteness of the $A$-atom density operator.
This supposition was somewhat strengthened by two additional
facts.

The general structure of the collision kernels:
\eqref{sjk1} In the ME approach and \eqref{sjsker} in the
standard one, is quite the same. The only external difference
consists in the structure of the "state selective" delta-like
factors. At the first sight it is not at all clear whether
these factors lead to the same "state selection" mechanisms.

Secondly, the collision rate $\Gamma$ is given in
Eq. \eqref{sjsgam} by the difference of forward scattering
amplitudes, while $\gamma_{ba}$ defined in \eqref{sjg1} clearly
contains products of scattering amplitudes, as it follows after
insertion of \eqref{sjk1} into \eqref{sjg1}.

These arguments seem to support the supposition that
the standard derivation is not certain to preserve
the positivity of the $A$-atom density operator. To clarify
these points we shall consider the $A$-atom with
multilevel but nondegenerate structure. Hence, we once again
take the free atom hamiltonian as: $H_{A} = \sum_{k}%
\hbar \omega_{k}\xket{k}\xbra{k}$, with kets $\xket{k}$
constituting an orthonormal and complete basis  in the Hilbert
space of atomic states. The free evolution of the elements
of the atomic density operator is given as
\begin{equation}
   \pdt \: \rho_{mn}(\wek{v}) \Bigr|_{free}
    = - i \omega_{mn} \rho_{mn}(\wek{v}),
\label{sjxx02} \end{equation}
while we assume that Bohr frequencies $\omega_{mn} \neq %
\omega_{jk}$ for different pairs of indices.

%%%-----------------------------------------------------------------
\subsection{Master equation approach}

The general structure of the master equation is given in
Eq. \eqref{sjme} and it must now be adapted to the presently
considered model of multilevel nondegenerate atom. The choice
of the operator basis is in this case obvious. We simply take
\begin{equation}
   S_{a} ~\longleftrightarrow~ P_{jk} = \xket{j} \xbra{k}.
\label{sjbase} \end{equation}
Hence, index $a$ used previously to enumerate the operator basis
is now replaced by a pair of numbers $(j,k)$. Moreover, Bohr
frequency $\Omega_{a}$ corresponds now to $\omega_{jk}$.
The considered density operator can then be expanded in the
chosen basis as
\begin{equation}
   \rho(\wek{v}) = \sum_{j,k} \rho_{jk}(\wek{v}) P_{jk}.
\label{sjro} \end{equation}
Using the adopted identifications all the terms in Eq.\eqref{sjme}
can easily be computed. Since we are mainly interested in the
comparison of the two variants of the collisional terms
of BBE we shall omit the computational technicalities.
The kernels ${\cal K}_{ab} = {\cal K}_{jk,mn}$
(and consequently the rates $\gamma_{ab} = \gamma_{jk,mn}$)
contain the "state selective" factors $\delta_{\Omega_{a},%
\Omega_{b}} =\delta(\omega_{jk} -\omega_{mn})$ acting as
Kronecker deltas. Careful but simple computation of all
the necessary sums leads to the following collisional
equations of motion: for populations we get
\begin{align}
   \pdt \: \rho_{mm}(\wek{v}) &   \Bigr|_{coll.}
   = - \widetilde{\gamma}_{mm}(\wek{v}) \rho_{mm}(\wek{v})
\nonumber \\
     & + \sum_k  \int d \wek{v}^{\:'}
         ~\mathcal{K}_{mk,mk}(\wek{v}) \rho_{kk}(\wek{v}^{\:'}),
\label{sjmepp} \end{align}
where we notice that the presence of the summation reflects
the fact that inelastic collisions are also accounted for.
On the other hand, for coherences we get
\begin{align}
   \pdt \: & \rho_{mn}(\wek{v}) \Bigr|_{coll.}^{(m \neq n)} =
\nonumber \\
   = &- \frac{1}{2}  \bigl[ \, \widetilde{\gamma}_{mm}(\wek{v})
        + \widetilde{\gamma}_{nn}(\wek{v})\,\bigr]
          \rho_{mn}(\wek{v}^{\:'})
\nonumber \\
     &+ \int d\wek{v}^{\:'}
       ~\mathcal{K}_{mm,nn}(\wek{v} \gets \wek{v}\,')
        \rho_{mn}(\wek{v}^{\:'}).
\label{sjmecc} \end{align}
In the two just given equations we have introduced a convenient
abbreviation
\begin{align}
  \widetilde{\gamma}_{mm} \bigl( \, \wek{v}_{1} \, \bigr)
  & = \sum_j \gamma_{jm, jm} \bigl( \, \wek{v}_{1} \, \bigr)
\nonumber \\
  & = \sum_{j} \int d\wek{v}
     ~\mathcal{K}_{jm,jm}(\wek{v} \gets \wek{v}_{1}).
\label{sjmg2} \end{align}
The collision kernel appearing here is now written in the
following form
\begin{align}
  \mathcal{K}_{mj, nk}(\wek{v} & \leftarrow \wek{v}_{1}) =
\nonumber \\
   = \; &  2 N_{P} \int d\wek{v}_{r} \int d\wek{v}_{r_1}
         ~W^{(P)}(\wek{v}_{1} - \wek{v}_{r_1})
\nonumber \\
   & \times
     \delta^3 \left[ \wek{v} - \wek{v}_{1}
            -\frac{\mu}{m_{a}} \left(\wek{v}_{r} - \wek{v}_{r_1}
     \right) \right]
\nonumber \\
   & \times
     \delta \left( v_{r}^{2} - v_{r_1}^{2}
    + \frac{2\hbar\omega_{mj}}{\mu} \right)
\nonumber \\
   & \times
     f(m,\wek{v}_r \gets j,\wek{v}_{r_1} )
    ~f^{*}(n,\wek{v}_r \gets k, \wek{v}_{r_1}),
\label{sjmek2} \end{align}
Once again we feel it necessary to stress that the resulting
equations \eqref{sjmepp}--\eqref{sjmecc} preserve hermiticity,
normalization and positivity of the atomic density
operator $\rho(\wek{v})$.

%%%-----------------------------------------------------------------
\subsection{Application of the standard approach}

Equation \eqref{sjse} has now to be transformed to suit the
currently investigated model. General (multi)indices $\alpha,%
\alpha\,'$ should be replaced by numbers $j,~k$, etc.
The corresponding changes are then to be made in the expressions
\eqref{sjsgam} and \eqref{sjsker} where the summations involving
the "state selective" delta-like factors can now be easily
performed. This leads to the following equation of motion
for populations
\begin{align}
   \pdt \: & \rho_{mm}(\wek{v}) \Bigr|_{coll.}
   = - \Gamma_{mm}(\wek{v}) \rho_{mm}(\wek{v})
\nonumber \\
     &+ \sum_{k} \int d\wek{v}_{1}
      ~\mathcal{J} \bigl(\,mm,\wek{v}\big|kk,\wek{v}_1\,\bigr)
       \rho_{kk}(\wek{v}_{1}),
\label{sjspp} \end{align}
where we again see the contributions from inelastic collisions.
The corresponding equation for coherences reads
\begin{align}
   \pdt \: & \rho_{mn}(\wek{v}) \Bigr|_{coll.}^{(m \neq n)}
   = - \Gamma_{mn}(\wek{v}) \rho_{mn}(\wek{v})
\nonumber \\
     & + \int d\wek{v}_{1}
    ~\mathcal{J} \bigl(\,mn,\wek{v}\big|mn,\wek{v}_1\,\bigr)
     \rho_{mn}(\wek{v}_{1}).
\label{sjscc} \end{align}
The collision rate \eqref{sjsgam} transformed to suit the
presently considered model is now given as
\begin{align}
  \Gamma_{mn}(\wek{v})
  \equiv & \; \Gamma\bigl(mn, \wek{v} \bigl| mn  \bigr)
\nonumber \\
  = & \; N_{P} \left( \frac{2 \pi \hbar}{i \: \mu} \right)
    \int d\wek{v}_{r}  ~W^{(P)} ( \wek{v} - \wek{v}_{r} )
\nonumber \\
    & \times \bigl[ f(m, \wek{v}_{r} \leftarrow m, \wek{v}_{r})
       - f^{\ast}(n, \wek{v}_{r} \leftarrow n, \wek{v}_{r}) \bigr].
\label{sjsgmn} \end{align}
and the collision kernel is of the following form
\begin{align}
   \mathcal{J} \bigl(mn,\wek{v} \big|& jk, \wek{v}_1 \bigr)
   = 2 N_{P} \int d \wek{v}_r  \int  d \wek{v}_{r1}
      ~W^{(P)} (\wek{v}_{1} - \wek{v}_{r1}\bigr)
\nonumber \\
   & \times \delta \left(\wek{v} - \wek{v}_1
       - \frac{\mu}{m_a}(\wek{v}_r - \wek{v}_{r1}) \right)
\nonumber \\
& \times
   ~\delta \left( v_{r}^{2} - v^{2}_{r1}
      + \frac{2}{\mu} ( \, E_m - E_j \, ) \, \right).
\nonumber \\
   & \times
   f(m,\wek{v}_r \gets j,\wek{v}_{r1}) \;
   f^{*}(n,\wek{v}_r \gets k,\wek{v}_{r1}).
\label{sjsk2} \end{align}

Definition \eqref{sjsgmn} implies that
$\Gamma_{mn}(\wek{v}) = \Gamma_{nm}^{*}(\wek{v})$.
Moreover, general relation \eqref{sjkast} yields
$\mathcal{J}^{\ast} \bigl(mn,\wek{v} \big| jk, \wek{v}_1 \bigr)%
=\mathcal{J} \bigl(nm,\wek{v} \big| kj, \wek{v}_1 \bigr)$.
These two facts ensure that Eqs. \eqref{sjspp} and \eqref{sjscc}
preserve hermiticity of the density operator.

To prove that the normalization is retained properly, we
need to show that relation \eqref{norm} is satisfied.
Due to Eq. \eqref{sjspp} is equivalent to the condition
\begin{equation}
   \Gamma_{kk}(\wek{v}_{1})
   = \sum_{m} \int d\wek{v}
    ~\mathcal{J} \bigl(\,mm,\wek{v} \big| kk,\wek{v}_{1}\,\bigr).
\label{sj:nn2} \end{equation}
Obviously, definition \eqref{sjsgmn} of the collision rate
implies that
\begin{align}
  \Gamma_{kk}(\wek{v})
   = N_{P} & \left( \frac{4 \pi \hbar}{ \mu} \right)
    \int d\wek{v}_{r} ~W^{(P)} ( \wek{v} - \wek{v}_{r} )
\nonumber \\
   &\times
    \mathrm{Im} \bigl\{ f(k,\wek{v}_{r} \leftarrow k,
    \wek{v}_{r}) \bigr\}.
\label{sj:nn3} \end{align}
Since the discussed formalism allows for inelastic scattering,
we need to use multichannel scattering theory \cite{taylor}.
Optical theorem allows us to express the imaginary part of the
elastic forward scattering amplitude by the total cross
section $\sigma_{T}( k,\wek{v}_{r})$ for scattering from
the state $\xket{k}, \wek{v}_{r}$. Thus, we cast the lhs of
\eqref{sj:nn2} into the form
\begin{equation}
  \Gamma_{kk}(\wek{v}_{1})
  = N_{P} \int d\wek{v}_{r}  ~W^{(P)} (\wek{v}_{1} - \wek{v}_{r})
    ~|\wek{v}_{r}|~\sigma_{T} \bigl( k,\wek{v}_{r} \bigr).
\label{sj:nn4} \end{equation}
The total cross section can be written as a sum
\begin{equation}
   \sigma_{T}(k,\wek{v}_{r})
   = \sum_{m} \int d\Omega(\wek{v}_{r1})
    ~\frac{d\sigma_{k \rightarrow m}}{d\Omega(\wek{v}_{r1})},
\label{sj:nn11} \end{equation}
where in the rhs we have differential cross sections
corresponding to scattering from state $\xket{k}, \wek{v}_{r}$
to $\xket{m}, \wek{v}_{r1}$ and where the integration is
performed over the angles specified by the direction of final
velocity. Then, the collision rate \eqref{sj:nn4} takes the
form
\begin{align}
  \Gamma_{kk}(\wek{v}_{1})
  = & N_{P} \sum_{m} \int d\wek{v}_{r}
    \int d\Omega(\wek{v}_{r1})
\nonumber \\
    & \times
    W^{(P)} (\wek{v}_{1} - \wek{v}_{r}) ~|\wek{v}_{r}|
    ~\frac{d\sigma_{k \rightarrow m}}{d\Omega(\wek{v}_{r1})}.
\label{sj:nn4x} \end{align}
On the other hand, rhs of Eq. \eqref{sj:nn2} contains square
moduli of scattering amplitudes (as it follows from
\eqref{sjsk2}). From multichannel scattering theory we have
\begin{equation}
   \bigl|f (m,\wek{v}_{r} \leftarrow k,\wek{v}_{r1}) \big|^{2}
   = \frac{|\wek{v}_{r1}|}{|\wek{v}_{r}|}
     ~ \frac{d\sigma_{k \rightarrow m}}{d\Omega(\wek{v}_{r})}.
\label{sj:nn5} \end{equation}
(note the reversed roles of relative velocities $\wek{v}_{r}$
and  $\wek{v}_{r1}$). Inserting the kernel \eqref{sjsk2} into
the rhs of \eqref{sj:nn2} and using \eqref{sj:nn5} we can
perform all the necessary integrations. Then, we arrive at the
expression identical with rhs of \eqref{sj:nn4x}.
This completes the proof of relation \eqref{sj:nn2} and,
therefore, we see that the considered model ensures
preservation of the proper normalization of $A$-atom density
operator. We note, however, that the question of positivity
preservation still remains open.

The general property \eqref{sjgpro} of the collisional rate
in standard approach allows us to write for the presently
considered case
\begin{equation}
   \mathrm{Re} \left\{ \: \Gamma_{mn}(\wek{v}) \: \right\}
    =  \frac{1}{2} \Bigl( \,
       \Gamma_{mm}(\wek{v}) + \Gamma_{nn}(\wek{v})\, \Bigr),
\label{sjgg} \end{equation}
which has important consequences. Eq. \eqref{sjscc}
describes the collisional evolution of coherences.
The term containing imaginary part of $\Gamma_{mn}$ can be
written separately. Then it can be combined with hamiltonian
(unitary) part of the evolution and identified as the
collisionally induced atomic frequency shift. Therefore,
only the term containing $\mathrm{Re} \left\{ \Gamma_{mn}%
\right\}$ contributes to relaxation part of Eq. \eqref{sjscc},
which is therefore replaced by the following equation
\begin{align}
   \pdt \: \rho_{mn}(\wek{v}) &  \Bigr|_{coll.}^{(m \neq n)}
   =  - \frac{1}{2}\Bigl(\Gamma_{mm}(\wek{v}) +
        \Gamma_{nn}(\wek{v})\Bigr) \rho_{mn}(\wek{v})
\nonumber \\
     & + \int d\wek{v}_{1}
      ~\mathcal{J} \bigl(\,mn,\wek{v}\big|mn,\wek{v}_1\,\bigr)
       \rho_{mn}(\wek{v}_{1}),
\label{sjat:cc} \end{align}
Moreover, we note that within the discussed model only the rates
$\Gamma_{kk}(\wek{v})$ given in \eqref{sj:nn2}-\eqref{sj:nn3}
or in \eqref{sj:nn4x} are of importance.

Summarizing, we can say that within the standard approach the
evolution of the density operator of a multilevel nondegenerate
atom is governed by Eqs.\eqref{sjspp} and \eqref{sjat:cc}
for populations and coherences, respectively. Finally, we note
that Rautian and Shalagin give corresponding equations which
do not fully agree with the above given results \cite{note01}.
We attribute this discrepancy most probably to  misprints.
Moreover, we have fully used the "state selective" delta-like
factors which, unfortunately, are not consequently resolved
by Rautian and Shalagin who retain exponential factors
like one in \eqref{sjexp}.

%%%-----------------------------------------------------------------
\subsection{Equivalence of both approaches}

Collisional equations of motion for both discussed approaches
are given by formulas \eqref{sjmepp} and \eqref{sjspp}
for populations, and by Eqs.  \eqref{sjmecc} and \eqref{sjat:cc}
for coherences, respectively. By inspection, we see that these
two pairs of equations have the same formal structure.
On the other hand, comparing Eqs. \eqref{sjmek2} and
\eqref{sjsk2} giving the kernels for respective approaches,
we see that they are, in fact, identical
\begin{equation}
  \mathcal{J} \bigl(mn,\wek{v} \big| jk, \wek{v}_1 \bigr)
  = \mathcal{K}_{mj, nk}(\wek{v}  \gets \wek{v}_{1}).
\label{sj:e01} \end{equation}
Next, from Eq. \eqref{sjmg2} it follows that
\begin{align}
  \widetilde{\gamma}_{mm} \bigl( \, \wek{v}_{1} \, \bigr)
  & = \sum_{j} \int d\wek{v}_{1}
     ~\mathcal{K}_{jm,jm}(\wek{v}_{1} \gets \wek{v})
\nonumber \\
  & = \sum_{j} \int d\wek{v}_{1}
     ~\mathcal{J}\bigl(jj,\wek{v}_{1} \big| mm, \wek{v} \bigr)
\nonumber \\
  & = \Gamma_{kk}(\wek{v}_{1}),
\label{sj:e02} \end{align}
where, in the last step, we have used relation \eqref{sj:nn2}
which was proved for the standard approach.

We conclude, that the pairs of equations
\eqref{sjmepp}--\eqref{sjspp} and
\eqref{sjmecc}--\eqref{sjat:cc} are not only of the same formal
structure but are strictly identical.
This allows us to answer the question on the preservation
of positivity of the atomic density operator within
the standard approach. Since master equation technique is
guaranteed to do so and the standard approach yields the same
results, it also preserves all the necessary properties of the
atomic density operator. Obviously, this conclusion is valid
for the considered model -- a multilevel atom with nondegenerate
levels satisfying the requirement that $\omega_{jk} \neq %
\omega_{mn}$ for two different pairs of indices.

It is perhaps worth noting that proving the equivalence of
both approaches we have used Eq. \eqref{sjat:cc} for evolution
of coherences in the standard approach. Certainly, this
equation is valid due to the possibility of replacing
$\Gamma_{mn}$ in Eq. \eqref{sjscc} by the right hand side
of Eq. \eqref{sjgg}. Hence, the latter one is essential in
the proof of equivalence and it follows directly from
definition \eqref{sjsgmn}, see also \cite{note01}).

%%%-----------------------------------------------------------------
\section{Final remarks \label{sec:fr}}

We have compared two different approaches to the derivation
of the collisional (Boltzmann) part of the spectroscopically
important Bloch-Boltzmann equations: the master equation
approach \cite{ak} and the standard approach as reviewed by
Rautian and Shalagin \cite{rash}. We have shown that both
approaches for a multilevel atom with nondegenerate levels
and with nondegenerate Bohr frequencies are equivalent.
Within the given model both preserve the fundamental
properties od the atomic density operator: hermiticity,
normalization and positivity. We have thus proved that the
supposition stated at the beginning of Sec.\ref{sec:dd}
was wrong and both approaches, at least within the discussed
model, are equivalent.

As it is seen from our comments lifting any of our
assumptions may lead to different conclusions. First of all,
we note that allowing for overlapping line profiles, that is
for $\omega_{jk}$ not necessarily different from $\omega_{mn}$
(with $(j,k) \neq (m,n)$) may lead to other results. In the
above considered case the collisional equations of motion for
coherences \eqref{sjmecc} and similarly \eqref{sjat:cc}
connect the given coherence $\rho_{mn}$ only to itself.
This is clearly due to the assumption that $\omega_{jk}%
\neq \omega_{mn}$.
In the above presented model the populations (see Eqs.
\eqref{sjmepp} or \eqref{sjspp} are coupled to other
populations. This population transfer is obviously due to
inelastic collisions which induce excitation--deexcitation
processes. It may be expected that when the assumption
$\omega_{jk} \neq \omega_{mn}$ is not valid, the same would
happen to coherences -- inelastic collisions would induce
polarization transfer, that is would couple coherence
$\rho_{mn}$ to other coherences. However, details of such
couplings would certainly depend on the degeneracies between
different Bohr frequencies. It is, therefore, difficult to
give any predictions on the specific couplings.
Perhaps, the best way is to examine some concrete model with
the discussed degeneracies and thus exhibiting polarization
transfer. Then, the problem whether the standard approach to
BBE still retains all the required properties of the atomic
density operator needs to be reexamined. In this context
we may say that the master equation technique is advantageous.
Due to its rigorous mathematical background it will
certainly preserve all the necessary properties of
$\rho(\wek{r},\wek{v},t)$. Indicated problems are clearly
of interest and seem to be a good subject for further
investigations. The question of conservation of positivity
could be answered again and the advantages of both approaches
would be weighted again.

Secondly, it is important to remember that we have taken
the perturbers to be structureless. It is, however,
not an essential simplification. It is straightforward
to generalize our results to unpolarized perturbers, as it is
consequently done by Rautian and Shalagin.
Additional degree of freedom (that is $\beta$ -- an index
indicating an internal state of the perturber) will result
in additional summations in the expressions for the collision
kernels and in additional terms in the Dirac deltas responsible
for energy conservation. The "state selective" factors either
in the standard or in the master equation method would remain
unchanged. Moreover, it seems that the assumptions concerning
the perturbers are less constraining. The reason, having
purely physical but not mathematical background, seems to be
simple. In the majority of spectroscopically interesting
situations the noble gas serves as perturbers. The excitation
energy of noble gas atoms is usually beyond the region of
interesting energies of $A$-atom transitions.
So, the perturbers indeed act as structureless particles.
As we already mentioned, it does not seem difficult to
generalize our master equation approach to perturbers with
full internal structure (discarding even the assumption about
nonpolarizability). The question is whether it is experimentally
interesting or relevant. But on the other hand, the problem
of positivity conservation within the standard approach
would need to be reexamined again.

Rautian and Shalagin introduce the "state selection" via the
exponential phase factors \eqref{sjexp}. They also express
their reservations about the validity of such an approach.
It seems that, at least within the model discussed in this work,
their method is equivalent to the secular approximation inherent
in the master equation. The proved equivalence of two
approaches seems to clarify the relationship between two
derivations of collisional terms in the BBE and to strengthen
the arguments given by Rautian and Shalagin. However,
when perturbers are allowed to be polarizable the phase factors
of Rautian and Shalagin do not necessarily coincide with the
delta factors resulting from secular approximation in the
master equation approach. Consequences of this, also present
another problem which deserves further investigations.

We hope that our work is useful to clarify some questions
concerning the various approaches to the derivation of
Boltzmann parts of BBE. We also hope that the results given
here would be a useful starting point for some further research.

%==================================================================
\begin{acknowledgments}
We would like to thank Professor Robert Alicki for comments
and discussion.
\end{acknowledgments}
%------------------------------------------------------------------

%%%================================================================

%%%-----------------------------------------------------------------

%%%=================================================================

\begin{thebibliography}{99}
\bibitem{note01}
    Namely, their collision rate (see \cite{rash}, p. 46,
    Eq. (2.157)) is written as follows
    \begin{align}
    \Gamma_{mn}^{(RS)}(\wek{v})
    & = N_{P} \left( \frac{2 \pi \hbar}{i \: \mu} \right)
        \int d\wek{v}_{r}~W^{(P)}(\wek{v}-\wek{v}_{r})
    \nonumber \\
    & \hspace*{8mm} \times
        \bigl[ f(m, \wek{v}_{r} \gets n, \wek{v}_{r})
         - f^{\ast}(m, \wek{v}_{r} \gets n, \wek{v}_{r}) \bigr].
    \nonumber \\
    & = N_{P} \left( \frac{4 \pi \hbar}{\mu} \right)
        \int d\wek{v}_{r}~W^{(P)}(\wek{v}-\wek{v}_{r})
    \nonumber \\
    & \hspace*{8mm} \times
        \mathrm{Im} \bigl[ f(m, \wek{v}_{r}
        \gets n, \wek{v}_{r}) \bigr],
    %\label{nn01}
    \nonumber \end{align}
    with notation changed to one used in this work.
    We note that the rate $ \Gamma_{mn}^{(RS)}$
    given by Rautian and Shalagin is, for populations,
    the same as our one given in Eq. \eqref{sj:nn3}.
    However, for coherences (when $m \neq n$) the rate
    $\Gamma_{mn}^{(RS)}$ does not agree with our relation
    \eqref{sjsgmn}. Hence, collision rate as given by
    Rautian and Shalagin does not have the property
    \eqref{sjgg}. This relation, in turn, is essential
    in our further discussion. We conclude, that the results
    of Rautian and Shalagin must be treated with care due
    to quite probable misprints. On the other hand, definition
    \eqref{sjsgmn} presented in this work agrees with
    expressions used by other authors, see for example
    \cite{hube}, Eq. 2.4.
\end{thebibliography}
\end{document}